\documentclass[12pt]{iopart}
\usepackage{iopams, graphicx}
\begin{document}
\title[SQUID-based multichannel system for Magnetoencephalography]{SQUID-based multichannel system for Magnetoencephalography}
\author{S Rombetto$^1$ C Granata$^1$ A Vettoliere$^1$ A Trebeschi$^2$ R Rossi$^2$ and M Russo$^1$}
\address{$^1$ Istituto di Cibernetica ``E. Caianiello" - CNR - Pozzuoli (Naples) - Italy}
\address{$^2$ Advanced Technologies Biomagnetics Srl, I-65129 Pescara, Italy}
\ead{s.rombetto@cib.na.cnr.it}
\begin{abstract}
Here we present a multichannel system based on  superconducting
quantum interference devices (SQUIDs) for magnetoencephalography
(MEG) measurements, developed and installed  at Istituto di
Cibernetica (ICIB) in Naples. This MEG system, consists of 163 full
integrated SQUID magnetometers, 154 channels and 9 references, and
has been designed to meet specifications concerning noise, dynamic
range, slew rate and linearity through optimized design.
The control electronics is located at room temperature and all the operations
are performed inside a Magnetically Shielded Room (MSR). The system
exhibits a magnetic white noise level of approximatively 5 fT/Hz$^{1/2}$.
This MEG system  will be employed for both clinical and routine use.
\end{abstract}
\pacs{74.81.Fa, 85.25.Hv,  07.20.Mc,  85.25.Dq, 87.19.le,  87.85.Ng}
\vspace{2pc}
\noindent{\it SQUIDs, Magnetoencephalography (MEG), superconductivity}
\maketitle
\section{Introduction}
Magnetoencephalography (MEG) is a non-invasive functional imaging
technique that measures the magnetic fields generated by neuronal
activity of the brain using Superconducting Quantum Inference
Devices (SQUIDs). Among the available brain imaging methods, MEG
uniquely features both a good spatial and an excellent temporal
resolution, thus allowing the investigation of key questions in
neuroscience and neurophysiology
\protect\cite{Ahonen1991}, \protect\cite{Johnson1996}, \protect\cite{Hari1993}, \protect\cite{Vrba1993}, \protect\cite{MEG}.\\
MEG measurements reflect intracellular electric current flow in the
brain and so they provide direct information about the dynamics of
evoked and spontaneous neural activity. Unlike  electroencephalogram
(EEG),  MEG is not subject to interferences due to the tissues and
fluids lying
between the cortex and the scalp \protect\cite{CohenMEGEEG}, \protect\cite{MEGEEG}.
As magnetic fields are not distorted by the different
conduction of the skull, MEG is an excellent localization tool for
subcortical sources of brain activity. It is worth noting that
measured magnetic fields are due exclusively to electric current
components tangential to the skull surface. Therefore the measured
signals are generated mainly in the cerebral sulci and there are only
minimal contributions due to neurons that present a different orientation.\\
MEG is a useful tool to investigate both
spontaneous  and evoked activities. MEG can be employed to investigate
the dynamic neuronal processes as
well as to study cognitive processes such as language
perception, memory encoding and retrieval and higher level tasks.
Moreover concerning  clinical
applications, it has been proved that MEG is a useful diagnostic
tool in the identification, prevention and treatment of numerous
disease and illnesses as it allows to study various cerebral
functions. So MEG has different applications, as for example
stroke, epileptic spike localization, presurgical functional
mapping, and Alzheimer disease.
\section{Multichannel system}
\begin{figure}
 \begin{minipage}{.5\textwidth}
 \includegraphics[bb = 0 0 198 265]{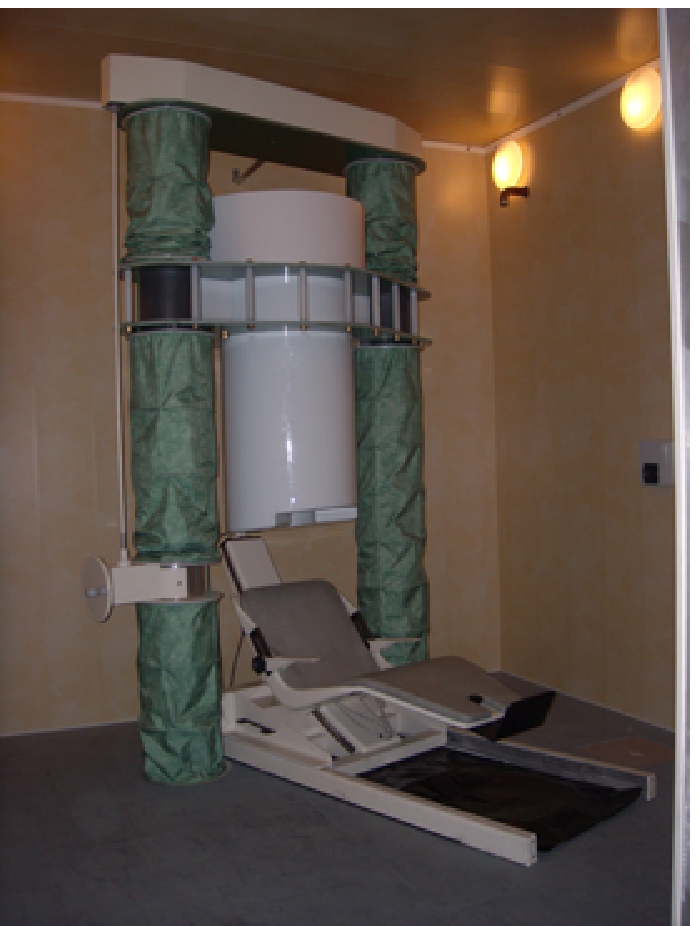}
\protect\caption{In the figure the whole MEG system, realized at Istituto di
Cibernetica, inside the MSR is shown. In particular the dewar
realized in fiberglass is visible.} \label{MEGdevice}
 \end{minipage}
 \begin{minipage}{.5\textwidth}
\includegraphics[bb = 0 0 227 170]{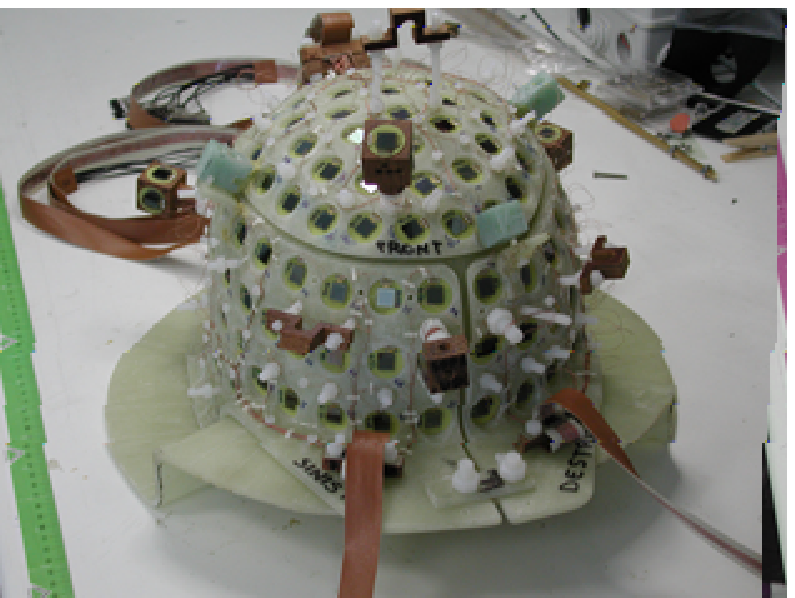}
\protect\caption{Helmet shaped array consisting
of 163 full integrated SQUID magnetometers, realized at ICIB-CNR.
The three reference triplets (consisting of 9 SQUIDs), are
visible.}\label{helmet}
 \end{minipage}
\end{figure}

Multichannel MEG systems take advantage of improvements in SQUID
technology, such as reproducibility, compact readout electronics,
and coil configuration. The MEG system described in this paper has
been developed at Istituto di Cibernetica in Naples in cooperation
with the Advanced Technologies Biomagnetics (AtB) and it is located
in our laboratory inside a clinic.\\ This MEG  consists of 163 full
integrated  SQUID magnetometers. A helmet shaped array hosts 154
magnetometer SQUIDs, located in a suitable way to cover the whole
head of the subject, and 3 vector magnetometers (consisting of 9
SQUIDs), located 6 cm above the helmet, and used as references
(Fig.\ref{helmet}). The reference magnetometers are perfectly
orthogonal and oriented along the coordinate axes.
\\
\subsection{SQUIDs}
\begin{figure}[!h]
\centering
\includegraphics[bb = 0 0 255 247]{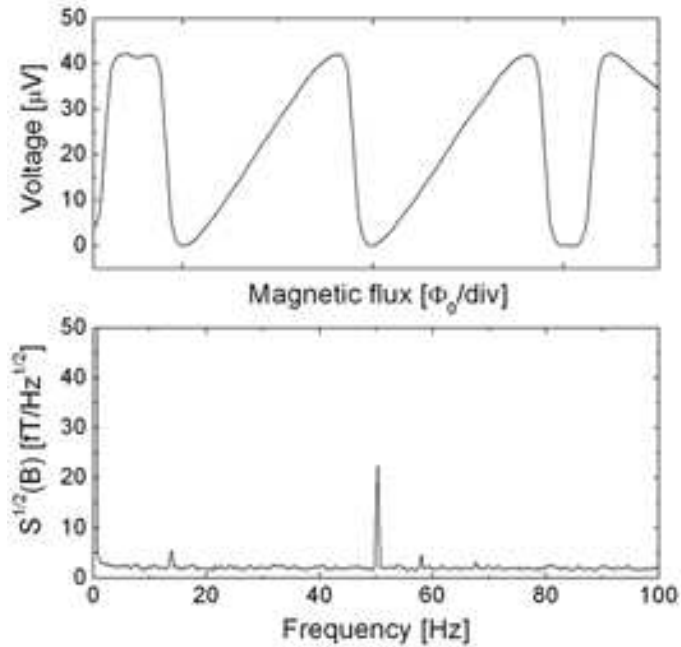}
\protect\caption{\label{singleSQUIDspectrum}
The voltage as a function of the magnetic flux (V-$\Phi$) for a
SQUID magnetometer and magnetic flux noise spectral density measured
at $T=4.2$ K are reported here.}
\end{figure}

SQUID magnetometers are very sensitive low frequency magnetic field
sensors  and present a spectral density of magnetic noise of a few
fT/Hz$^{1/2}$ \protect\cite{SQUIDbooka}, \protect\cite{SQUIDbookb}. Due to their
excellent performances, SQUID sensors are widely employed in several
applications ranging from biomagnetism, to magnetic microscopy, quantum computing, non
destructive evaluation test and geophysics \protect\cite{SQUIDbookb}. For
the MEG system described here SQUID sensors have been realized using
a standard trilayer technology, that ensures good performances
during time and a good signal to noise ratio, even at low
frequencies. Each SQUID magnetometer includes an integrated
superconducting flux transformer working as a magnetic flux pickup,
to increase the magnetic field sensitivity \protect\cite{Chieti_2001}.\\
The design of the sensor is based on a Ketchen-type magnetometer
\protect\cite{Ketchen}. The SQUID loop is a square planar washer with an
inductance of 260 pH, coupled to a 12-turn thin film input coil with
33 nH inductance connected in series with a square single turn
pickup coil of 64 mm$^2$ area presenting an inductance of 27 nH. The
Additional Positive Feedback (APF) \protect\cite{APF} circuit and the
feedback coil for Flux-Locked-Loop (FLL) operation \protect\cite{FLL} are
fully integrated on chip to avoid any additional noise due to an
external APF circuit. In order to obtain a high effective
flux-capture area of 3 mm$^2$, corresponding to a flux-field
conversion factor of 0.7 nT/$\Phi_0$, the mutual inductance between
the input coil and the SQUID has been
increased by a much higher SQUID inductance \protect\cite{Cantorbook}.\\ In
Fig.\ref{singleSQUIDspectrum} we report an experimental measurement of
the voltage as a function of the magnetic flux (V-$\Phi$) for a
SQUID magnetometer and magnetic flux noise spectral density measured
at $T=4.2$ K.
\\
The SQUID sensors are arranged in close proximity to each other over
the helmet shaped array reported in Fig.\ref{helmet}. This geometry
requires a particular care for both wire arrangement and design of
the SQUID support to minimize cross-talk \protect\cite{crosstalk}.
\subsection{Dewar}
For proper operation SQUIDs have to work at a temperature of 4.2 K,
reached immersing them in liquid helium. The dewar enclosing the
SQUIDs is an important component and must satisfy severe
requirements \protect\cite{Hari1993}.
In our system the dewar has been realized in fibreglass as this
material shows both excellent magnetic properties and optimal
thickness. This choice has allowed to minimize the distance between
head and  SQUIDs, so that the sensors are located only 2 cm
away from the scalp. Furthermore, to reduce the radiation losses,
several layers of mylar have been enclosed inside the inner portion
of the dewar. \\The dewar has a capacity of 74 liters and a helium refill
interval of 7 days, thanks also to a mold realized in foam that
minimizes the heat transfer, as described in section 4.
\subsection{Technical equipment}
\begin{figure}[!h] \center
\includegraphics[bb = 0 0 403 227]{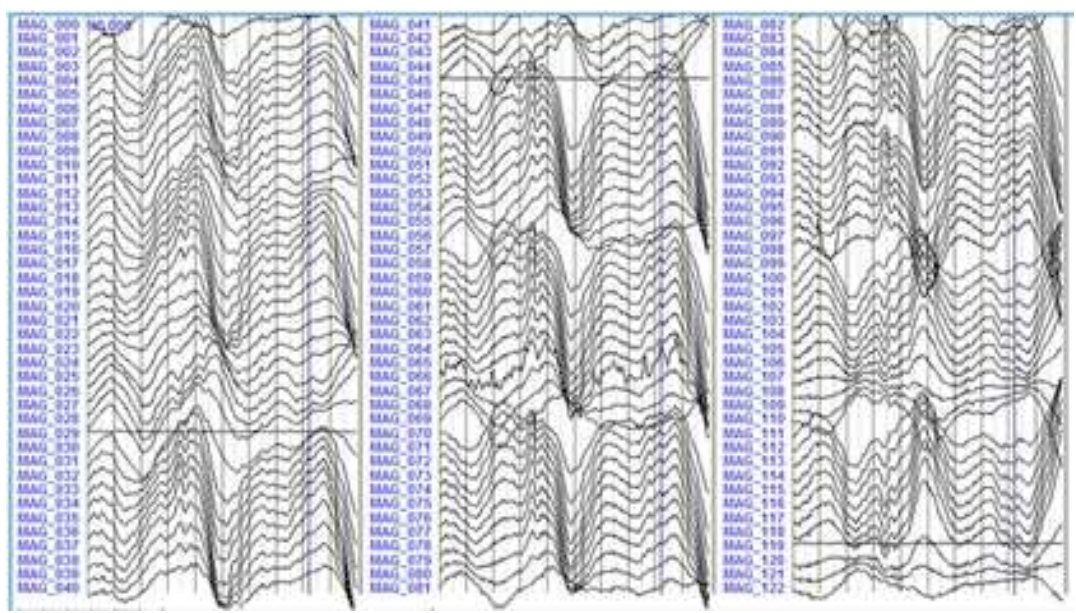}
\protect\caption{Here a sketch of preliminary measurement of the spontaneous activity
recorded by the MEG system on a healthy voluntary subject is shown.}
\label{caratteristicheSQUID}
\end{figure}

The readout electronics is placed at room temperature and is based
on the FLL  configuration \protect\cite{FLL} with direct coupling to the
preamplifier and an APF circuit \protect\cite{APF}. The contribution of
electronic noise due to preamplifier has been limited by increasing
the gain of the SQUIDs \protect\cite{APF}, \protect\cite{crosstalk}. Each SQUID is
connected to the room temperature electronics through four shielded
wires. Furthermore a suitable feedback coil integrated on the the
SQUID magnetometer chip, prevents the cross-talk phenomenon between
adjacent channels and allows the integration of a large number of
channels \protect\cite{crosstalk2}. In this configuration the SQUIDs are
directly coupled to the amplifiers at room temperature and work in
optimal conditions.\\
\begin{figure}[!h] \center
\includegraphics[bb = 0 0 340 182]{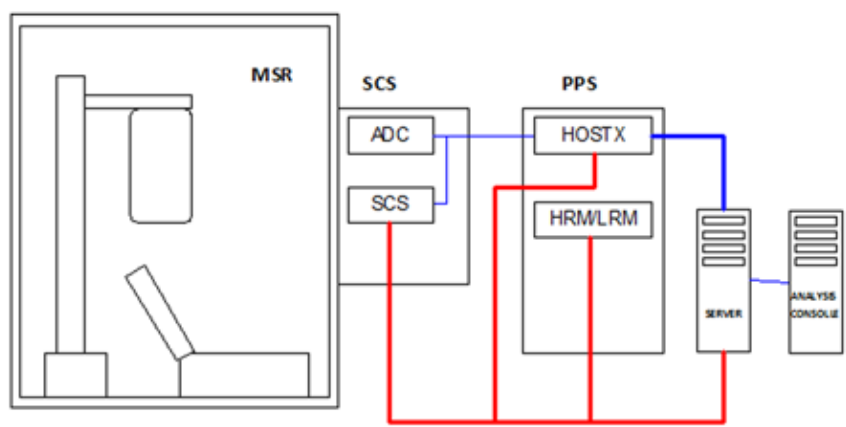}
\protect\caption{Block diagram of the MEG system at Istituto di Cibernetica.
SCS is the Sensor Control System, PPS is the Pre Processing System,
HRM and LRM are respectively the High Resolution
Monitor and the Low Resolution Monitor.} \label{BlockDiagram}
\end{figure}
\\Continuous data can be acquired simultaneously from
all channels in  the bandwidth DC-400
Hz. The measured signals are then A/D converted at a sampling
frequency of 8.2 kHz and sent to a group of digital signal
processors (DSP). The DSP group is controlled through the console
and is used to apply digital filters. Furthermore it guarantees that
all SQUID channels are sampled simultaneously at 1.025 kHz  with
22-bit ADCs (Analog to Digital Converter). SQUIDs' parameters can be
changed through the acquisition console by the operator.\\A block
diagram of our MEG system is shown in Fig.\ref{BlockDiagram}. SCS is the
Sensor Controller System, while the PPS is the Pre Processing
System, HRM and LRM are respectively the High Resolution Monitor and
the Low Resolution Monitor, HOSTX indicates computers that  contain
DSP cards.
\\Using filters
ensures that the digital signal processing is the same for each
channel, so that there are no delays between different channels and
that there are no jitters.\\In Fig.\ref{caratteristicheSQUID} a
preliminary measurement of the spontaneous activity
recorded by the MEG system on a healthy voluntary subject is shown.
\subsection{EEG system}
The MEG system is also equipped with a 32 integrated non-magnetic
EEG-channels cap, with ultra-thin wires and low profile of
electrodes, that is optimal for usage inside a MEG helmet
\protect\cite{ASA1}, \protect\cite{ASA2} and sintered Ag/AgCl electrodes guarantee an
optimal EEG signal quality and do not need  for re-chloriding.
Data are transmitted at 24-bit resolution via optical cable to
USB. This system allows to digital store the data and analyze them
by using both commercial and open source programs.
\\
\\
There is also the possibility to record  EOG and ECG using
additional electrodes. Scalp EEG can be inspected visually in
real time.
\section{MSR}
\begin{figure}[!h]
\centering
\includegraphics[bb=0 0 283 302]{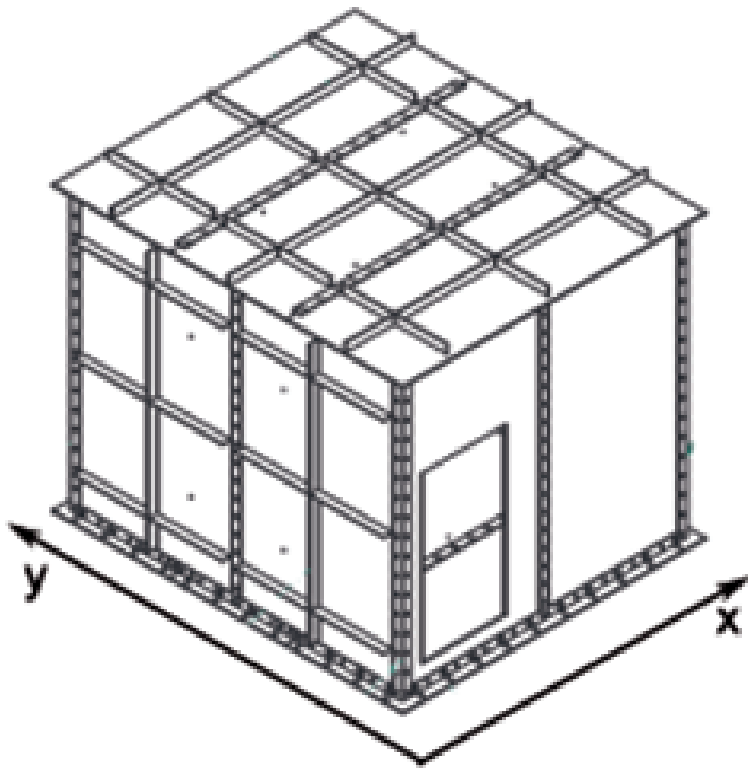} \protect\caption{\label{cabinArrow}
Here we report a sketch of the Magnetic Shielded Room containing MEG
system.}
\end{figure}

Since the magnetic signals emitted by the brain are about eight-nine
orders of magnitude smaller than the magnetic disturbances arising
from earth magnetic field and urban noise, it is necessary to shield
from external magnetic signals. The most straightforward noise
reduction method is  to  place the MEG system within a Magnetically
Shielded Room (MSR), that allows to physically reduce the
environmental noise. \\
A block diagram of the MEG system is illustrated in
Fig.\ref{BlockDiagram} while a sketch of our system is reported in
Fig.\ref{cabinArrow}.
\\
The MSR has been realized in aluminium and $\mu$-metal to reduce
respectively high-frequency and low-frequency noise. Our MSR
consists of three nested main layers: a pure aluminum layer (1.5 cm)
and  two $\mu$-metal layers (1.5 mm). Magnetic continuity is
maintained by overlay strips. The external dimensions of our MRS are
$3.7\times 4.3\times 3.4$ m$^3$ (l $\times$ w $\times$ h), and the
inner dimensions are $2.9\times 3.7\times 2.9$ m$^3$. All the
electric connections have been designed to not introduce any
magnetic noise. The floor is  independently suspended.\\
\\
The low  noise dewar and the MRS have allowed us to obtain a white
noise level of a white noise spectrum of approximately 5
fT/Hz$^{1/2}$  that is good if
compared to the average intrinsic noise of SQUIDs realized at
ICIB-CNR, that is around 2 fT/Hz$^{1/2}$ (Fig.\ref{singleSQUIDspectrum}).
\subsection{MSR attenuation} 
\begin{figure}[!h]
\centering
\includegraphics[bb = 0 0 457 340]{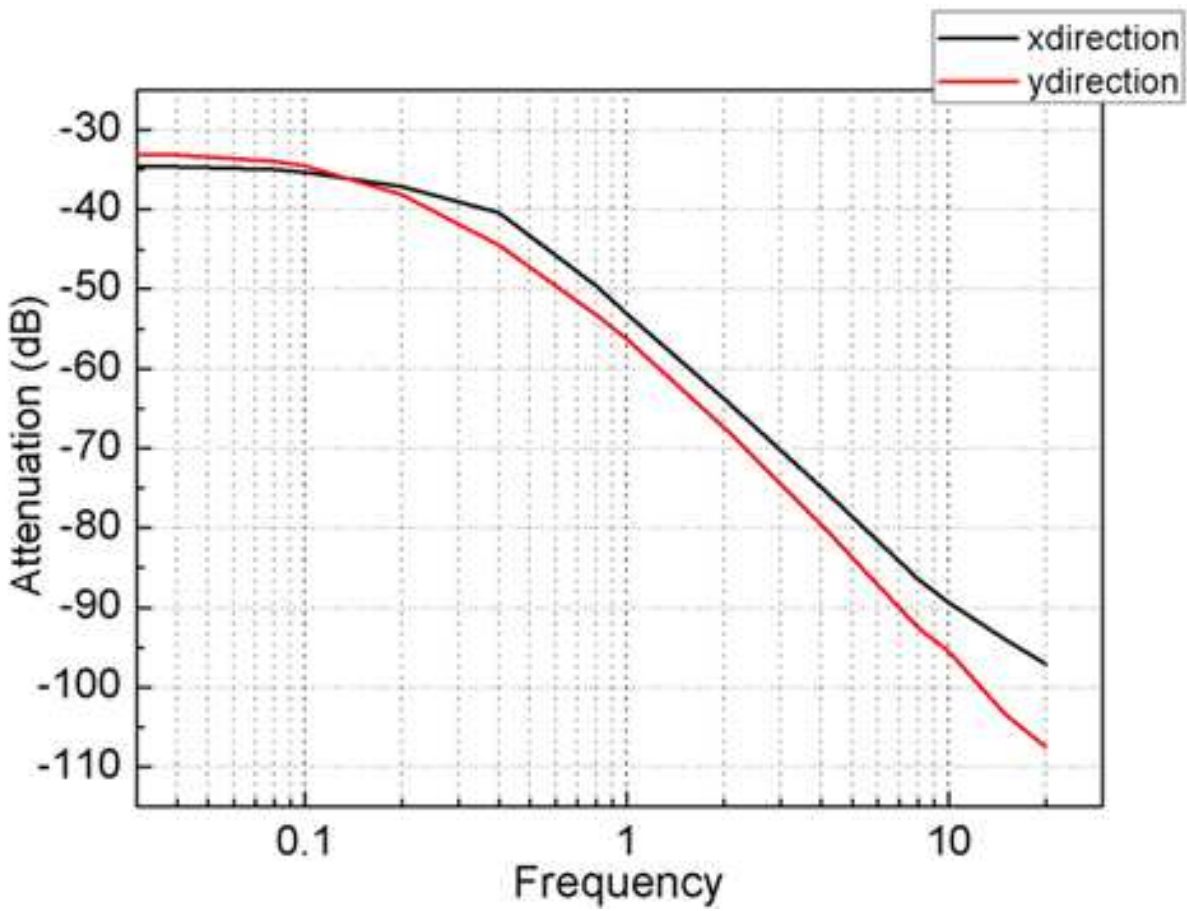}
\protect\caption{\label{attenuazione} The magnetic shielding factor of the
Magnetically Shielded Room as a function of the frequency for x
(black line) and y (red line) direction.}
\end{figure}

Here we report the measurements
made to characterize the performance of the shielded chamber. We
have measured  the response of MSR to an external applied field
produced by a pair of coils for  the \emph{x} and \emph{y}
component, in the frequency
range 0.01 Hz-20 Hz  (Fig.\ref{attenuazione}).
Passive shielding factors of 17266 and 2200 at 0,1 Hz and 1 Hz
respectively have been reported. Results are summarized in Table
\ref{tableMSR}.\\
We have observed that when a triangular waveform is injected inside
the coils, the  output of SQUIDs is a sinusoidal waveform. This
behaviour is explained considering that MSR filters high frequency
components. It has been observed also that the attenuation in the
\emph{y} direction is slightly lower than in \emph{x}
direction, probably due to the presence of the door.
\begin{table}[h!]
\center
\begin{tabular}{|c|c|c|c|}
\hline Frequency  & Amplitude  & Attenuation x  & Attenuation y \\
       (Hz)       & (pT)       & (dB)           & (dB)\\
\hline
\hline 0,01  & 18700 & -34,56  & -32,80\\
\hline 0,02  & 18900 & -34,47  & -32,99\\
\hline 0,04  & 18533 & -34,64  & -33,17\\
\hline 0,08  & 17900 & -34,94  & -33,96\\
\hline 0,1   & 17266 & -35,26  & -34,56\\
\hline 0,2   & 14000 & -37,08  & -38,20 \\
\hline 0,4   & 9500  & -40,45  & -44,46 \\
\hline 0,8   & 3300  & -49,63  & -53,19 \\
\hline 1     & 2200  & -53,15  & -56,29 \\
\hline 2     & 650   & -63,74  & -67,34 \\
\hline 4     & 180   & -74,89  & -79,48 \\
\hline 8     & 48    & -86,38  & -92,55 \\
\hline 10    & 34    & -89,37  & -95,47 \\
\hline 15    & 20    & -93,98  & -103,43 \\
\hline 20    & 14    & -97,08  & -107,51 \\
\hline
\end{tabular}
\protect\caption{Measured attenuation of the MSR to an external applied
field in the frequency range 0.01 Hz - 20 Hz, for the \emph{x} and
\emph{y} components. The amplitude is peak to peak.}\label{tableMSR}
\end{table}
\section{System thermogram}
We have used a thermal imaging camera to detect radiation in the
infrared range of the electromagnetic spectrum  and produce  thermal
images (thermograms) of the MEG system. In fact, as infrared
radiation is emitted by all objects above absolute zero according to
the black body radiation law, thermography allows to see and to
measure variations in temperature \protect\cite{Thermal}, \protect\cite{Rippa}.
\begin{figure}[!h]
\center
\includegraphics[bb=0 0 232 157]{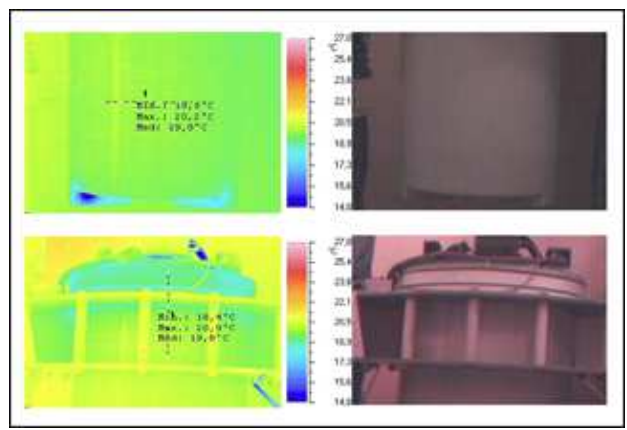}
\protect\caption{Thermograms (thermal images) of the MEG system taken using
a thermal imaging camera to monitor the thermal losses around the
dewar. The values displayed have been evaluated through a thermal
analysis of IR images.} \label{Termica}
\end{figure}\\
From the thermograms shown in Fig.\ref{Termica} it is evident that
this system has an excellent thermal resistance. Major variations are
observed on the joints and in the proximity of the helmet, due to
the reduced thickness of the dewar at those points. It is worth noting
that the temperature variations are within 1$^\circ$C. \\As a consequence
of these measures a mold in foam has been realized. When the system
is not performing measurements, it is placed inside the dewar, where
usually the head of the subject is placed, it allows to minimize the
thermal swapping.
\section{Spectral And Seismic Noise Analysis}
We have studied the Power Spectral Density (PSD) in the MEG
laboratory environment   by the means of seismic noise analysis
techniques \protect\cite{NotaInternaPaola}.
\\
Seismic instrumentation has been used to measure the effective
motion inside the MEG laboratory. In fact before installing a MEG
system is also necessary to perform  a careful analysis of seismic
noise as large mechanical vibrations can induce  a low-frequency
noise that affect the proper functioning of the SQUIDs.\\
\begin{figure}[!h]
\center
\includegraphics[bb = 0 0 397 320]{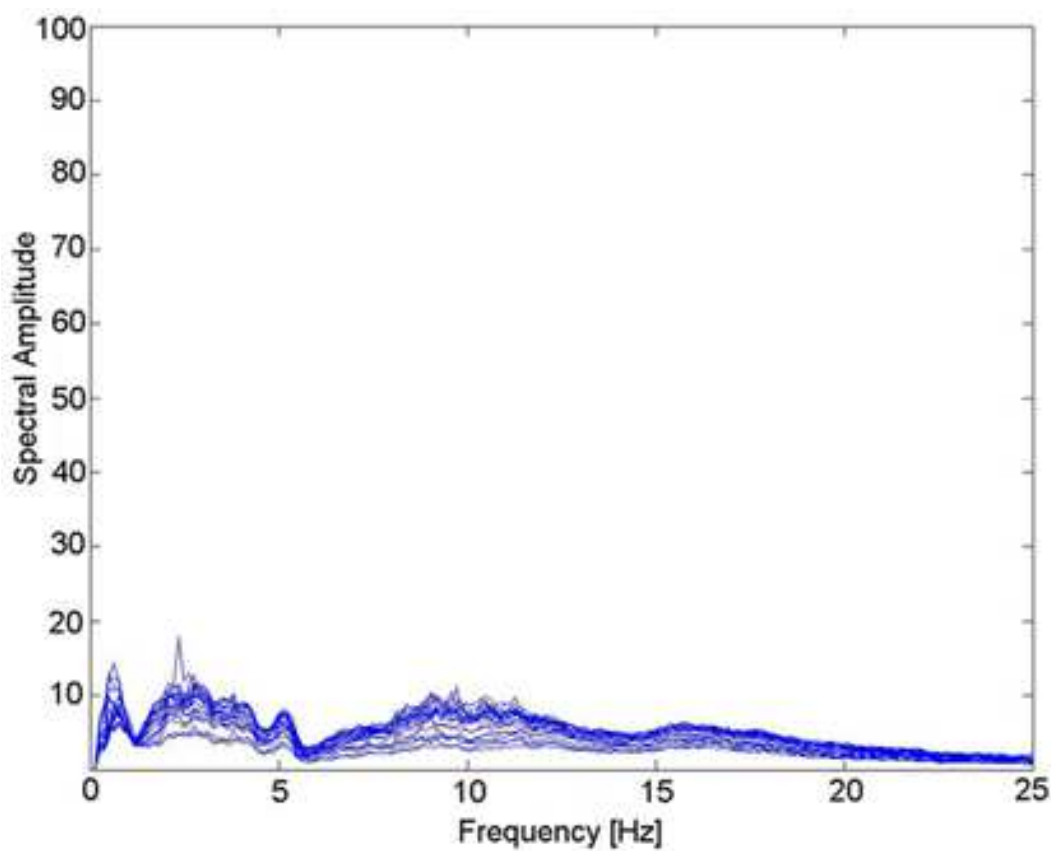}
\protect\caption{\label{spettriMEG}Power Spectral Density (PSD) of  the MEG
laboratory environment is shown.  Unit for the spectral amplitude is
($\mu$m/s)/Hz.}
\end{figure}\\
Data were collected along 4 days, including a week-end, to study the
noise in different conditions \protect\cite{NotaInternaPaola}. Data analysis
has shown that noise  inside the MEG laboratory is present
especially at really low frequencies, as shown in
Fig.\ref{spettriMEG}. We plan to introduce antivibration
pads below the MSR to avoid undesired mechanical oscillations and to
use noise reduction techniques, as for example active compensation
and off-line software methods.
\section{Conclusion}
We showed main characteristics, the technical equipment and the
performance of our  MEG system, consisting of 163 full integrated
SQUIDs developed at Istituto di Cibernetica and located in a
clinical environment.\\
The presented MEG system presents good characteristics for clinical and routine
use. The noise floor is about 5 fT/Hz$^{1/2}$ and  sensor performances are stable during operation.  This guarantees  that high-quality MEG recordings are possible with this system.
\section*{Acknowledgment}
This work was partially supported by Italian MiUR under the Project
''Sviluppo di componentistica superconduttrice avanzata e sua
applicazione a strumentazione biomedica" (L. 488/92, Cluster  14 -
Componentistica Avanzata).\\
\section*{References}
\bibliographystyle{unsrt}
\bibliography{manuscript}
\end{document}